\documentclass[%
 reprint,
superscriptaddress,
showpcas,
floatfix,
 aps,
 prl, 
]{revtex4-1}
\usepackage{booktabs}
\usepackage{upgreek}
\usepackage{makecell}
\usepackage{amsmath,amsfonts,amsthm}
\usepackage{amssymb}
\usepackage{graphicx}
\usepackage{dcolumn}
\usepackage{bm}
\usepackage{hyperref}
\hypersetup{colorlinks=true, citecolor=blue, urlcolor=blue, linkcolor=blue}

\usepackage{color}

\begin{document}

\preprint{APS/123-QED}

\title{
  Hall effect on the joint cascades of magnetic energy and helicity in helical magnetohydrodynamic turbulence  
}

\author{Running Hu}
\affiliation{%
  LHD, Institute of Mechanics, Chinese Academy of Sciences, Beijing 100190, PR China
}%
\affiliation{%
  School of Engineering Science, University of Chinese Academy of Sciences, Beijing 100049, PR China
}
\author{Jin-Han Xie}
\affiliation{%
Department of Mechanics and Engineering Science, College of Engineering, Peking University, Beijing 100871, PR China
}%
\affiliation{
  State Key Laboratory of Turbulence and Complex Systems, College of Engineering, Peking
University, Beijing 100871, PR China
}

\author{Xinliang Li}%
\affiliation{%
  LHD, Institute of Mechanics, Chinese Academy of Sciences, Beijing 100190, PR China
}%
\affiliation{%
  School of Engineering Science, University of Chinese Academy of Sciences, Beijing 100049, PR China
}
\author{Changping Yu}%
\email{cpyu@imech.ac.cn}
\affiliation{%
  LHD, Institute of Mechanics, Chinese Academy of Sciences, Beijing 100190, PR China
}%
\author{Yuan Hu}
\affiliation{%
  LHD, Institute of Mechanics, Chinese Academy of Sciences, Beijing 100190, PR China
}%
\author{Jianchun Wang}
\affiliation{%
Department of Mechanics and Aerospace Engineering, Southern University of Science and Technology, Shenzhen, Guangdong 518055, PR China
}%
\author{Shiyi Chen}
\affiliation{
  Eastern Institute for Advanced Study, Ningbo 315200,  PR China
}
\affiliation{
  State Key Laboratory of Turbulence and Complex Systems, College of Engineering, Peking
University, Beijing 100871, PR China
}
\affiliation{%
Department of Mechanics and Aerospace Engineering, Southern University of Science and Technology, Shenzhen, Guangdong 518055, PR China
}%

\date{\today}

\begin{abstract} 
  \par
  Helical magnetohydrodynamic turbulence with Hall effects is ubiquitous in 
heliophysics and plasma physics, such as star formation and solar activities, and its intrinsic mechanisms are still not clearly explained. Direct numerical simulations reveal that when the forcing scale is comparable to the ion inertial scale, Hall effects induce remarkable cross helicity. It then suppresses the inverse cascade efficiency, leading to the accumulation of large-scale magnetic energy and helicity. The process is accompanied by the breaking of current sheets via filaments along magnetic fields.  Using the Ulysses data, the numerical findings are separately confirmed. These results suggest a novel mechanism  wherein small-scale Hall effects could strongly affect large-scale magnetic fields through cross helicity.
 \end{abstract}

\maketitle

\setlength{\lineskip}{0.1pt}
\setlength{\lineskiplimit}{0.1pt}

\par 
In Hall magnetohydrodynamic (HMHD) turbulence, the  ions and electrons decouple from each other, which can be modeled via the Hall term in the generalized Ohm's law \citep{galtier2016Introduction}.
HMHD model is valid for the process with timescales shorter than the ion cyclotron period, such as star formation \citep{marchand2019Impacta,norman1985Anomalous,wurster2021Impact}, dynamo action \citep{gomez2010Hallmagnetohydrodynamic,mininni2005Direct}, planetary magnetosphere \citep{dorelli2015Role},  and solar activities \citep{bhattacharjee2004Impulsive,cassak2005Catastrophe}.
In the solar corona, the strongly helical magnetic field collisionlessly reconnects, leading to the rapid eruption of plasma \citep{ebrahimi2015,wan2015Intermittent,faganello2008Time,servidio2009Magnetic,sun2022Physical}.
\par In solar wind observation, the turbulent nature can be verified through the von K\'{a}rm\'{a}n-Howarth (vKH) relations  for magnetohydrodynamic (MHD) turbulence \citep{dekarman1938Statistical,politano1998Dynamical,marino2011Magnetohydrodynamic,marino2012Occurrence,sorriso-valvo2007Observation}.
Based on vKH relations, \citet{smith2009Turbulent} discovered prevailing inverse energy cascades when the cross helicity  is pronounced. 
\citet{marino2012Occurrence} found that cross helicity could affect the linear scaling ratio of the vKH relation.
In MHD and HMHD turbulence, magnetic helicity is another conservative quantity besides energy, and could measure the magnetic reconnection process quantitatively \citep{galtier2016Introduction}.
\citet{Politano2003Von} derived the  vKH relations for magnetic helicity in MHD turbulence.
\citet{banerjee2016Chiral} derived the balance for magnetic helicity in HMHD turbulence.
\citet{brandenburg2011Scale} found the  sign change of magnetic helicity at small- and large-scale solar wind.
\par In practice, even if solar wind can be described by MHD turbulence, the detailed mechanisms are hard to be investigated in actual space plasma due to its complex nature.
High-precision direct numerical simulations (DNS) are necessary. DNS of the helical MHD turbulence showed that helicity injection could lead to the formation of a large-scale magnetic field \citep{muller2012Inverse,richner2022Magnetic}.
\citet{mininni2005Direct} found the faster growth rates of large-scale magnetic fields in HMHD turbulence.
\citet{dreher2005Formation} studied the formation and disruption of Alfv\'{e}nic filaments under Hall effects.
\par However, even if there have been separate studies on magnetic helicity and Hall effects, there remains limited researches on their coupling effects, which are significant in the collisionless reconnection of the solar corona \citep{ebrahimi2015}.
The vKH relation about the magnetic helicity in HMHD turbulence  is still not addressed, and the existing related theoretical works are also not verified.
In this letter, we will investigate the Hall effects on the cascades of magnetic energy and helicity, and then verify those findings based on vKH relations in actual solar wind.
\par The governing equations for the velocity $\bf v$ and the magnetic field $\bf b$ are \citep{marino2023Scaling,muller2012Inverse}:
\begin{equation}
  \begin{aligned}
    \label{equ:control}
    \partial_t {\bf v} & =-({\bf v} \cdot {\bf\nabla} ){\bf v}
    -{\bf\nabla} p+{\bf j}\times {\bf b} +\lambda\Delta^{-1}{\bf v}+\nu \Delta {\bf v}+{\bf f_v},  \\
    \partial_t {\bf b} & =\nabla \times ({\bf U}\times {\bf b})
      +\lambda_b\Delta^{-1}{\bf b}+\eta \Delta {\bf b} +{\bf f_b},\\
  \end{aligned}
\end{equation}
where $\nabla \cdot {\bf v}=0$ and $\nabla \cdot {\bf b}=0$,  $p$ is the pressure, ${\bf U}={\bf v}-d_I{\bf j}$, $d_I$ is the ion inertial length, ${\bf j}=\nabla \times {\bf b}$ is the current density,  ${\bf f_v}$ and ${\bf f_b}$ are forcing terms at forcing wavenumbers $k_f$,    $\lambda\Delta^{-1} {\bf v}$ and $\lambda_b\Delta^{-1} {\bf b}$ are hypo-viscous terms, $\nu$ and $\eta$ are the kinematic viscosity and magnetic diffusivity, respectively.
\par Six DNSs are performed in this letter. 
The computational parameters are shown in TABLE~\ref{tab:dat}.
M3, M6 and M40 are the MHD cases with different $k_f$, while H3, H6 and H40 are corresponding HMHD cases.
The random forcing schemes (${\bf f_v}$ and ${\bf f_b}$) only inject maximum magnetic helicity artificially but do not inject any hydrodynamic or cross helicity.
See the Supplemental Material (SM) \citep{supplemental_material} for more numerical details.
\begin{table}[tb]
  \caption{\label{tab:dat}  The computational configurations  of  flows.
  $\varepsilon_v$ and $\varepsilon_b$ are injection rates of hydrodynamic and magnetic energy.
  }
  \begin{ruledtabular}
    \begin{tabular}{c|cccccccccc}
      Case & \makecell[c]{Grid} & \makecell[c]{$d_I$}  & $\nu=\eta$  & $\lambda=\lambda_b$ & \makecell[c]{$k_f$} & $\varepsilon_v=\varepsilon_b$  \\
      \colrule
      M3 & $2048^3$& -- & $4\times 10^{-5}$  & 0.1  & 3 & 0.015 \\
      H3 & $2048^3$& 0.05  &  $4\times 10^{-5}$  & 0.1  & 3 & 0.015 \\
      M6 & $2048^3$& -- & $4\times 10^{-5}$  & 0.1  & 6 & 0.015 \\
      H6 & $2048^3$& 0.05  &  $4\times 10^{-5}$  & 0.1  & 6 & 0.015 \\
      M40 & $1024^3$& --  & $10^{-4}$  & 0.05  & 40-42 & 0.015 \\
      H40 & $1024^3$& 0.05  &  $10^{-4}$  & 0.05  & 40-42 & 0.015 \\
    \end{tabular}
  \end{ruledtabular}
\end{table}
\begin{figure}[tb]
  \centering
  \includegraphics[width=\linewidth]{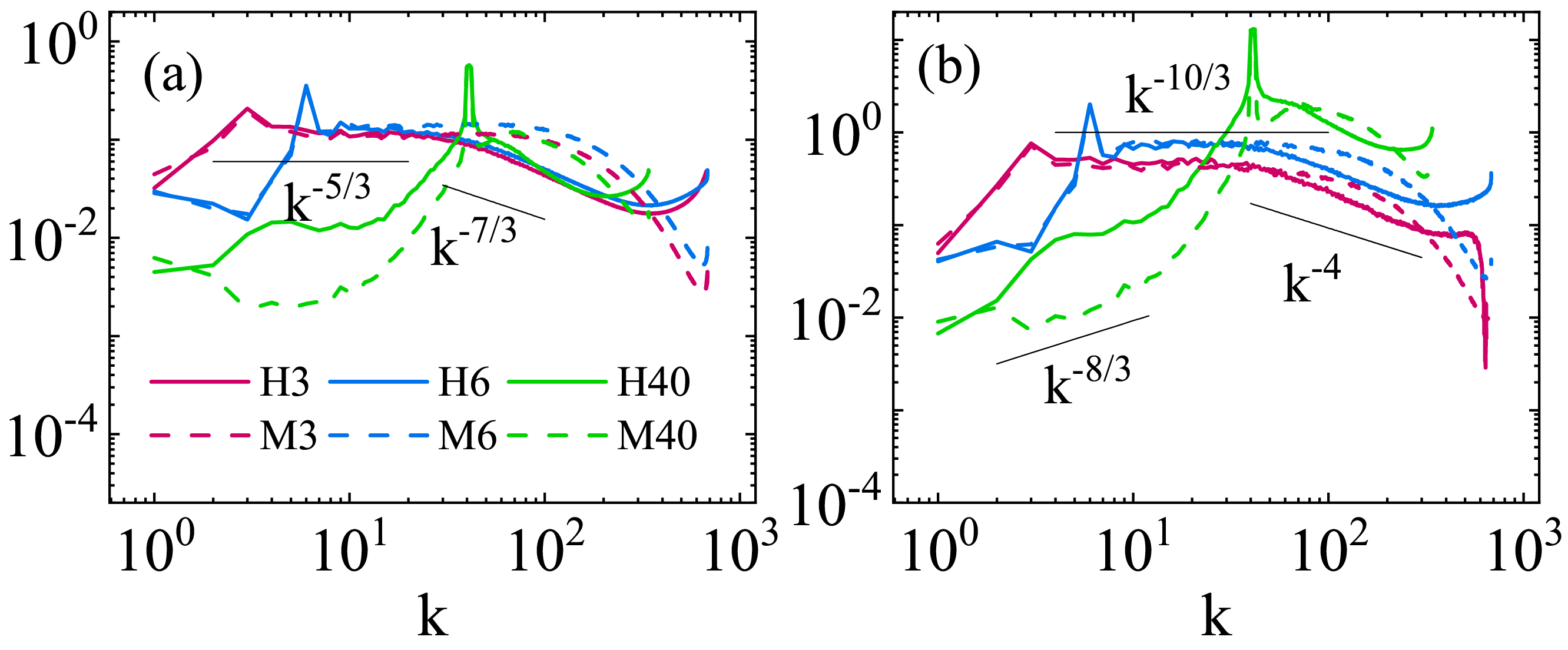}
  \caption{Compensated Spectra. 
  (a)  $E_b(k)k^{5/3}$.
  (b)  $H_M(k)k^{10/3}$.
  }
  \label{fig:spec}
\end{figure}
\par FIG.~\ref{fig:spec} shows the  spectra of magnetic energy $E_b(k)=\sum_{|{\bf k}|=k}\Re\{ \hat {\bf b}({\bf k})\cdot \hat {\bf b}^\ast({\bf k})\}/2$  and magnetic helicity $H_M(k)=\sum_{|{\bf k}|=k}\Re\{ \hat {\bf a}({\bf k})\cdot \hat {\bf b}^\ast({\bf k})\}$, where $\hat \cdot$ represents the values in spectral space, and $\hat {\bf a} ({\bf k})$ is the magnetic potential. 
In FIG.~\ref{fig:spec} (a), $E_b(k)\sim k^{-5/3}$ in the inertial range ($k\in (4,100)$), but in the ion inertial scale ($k \sim 1/d_I$), $E_b(k)\sim k^{-7/3}$ for HMHD cases (H3 and H6).
For the magnetic helicity in FIG.~\ref{fig:spec} (b), $H_M(k)\sim k^{-10/3}$ for  M3 and M6, due to the Alfv\'{e}nic balance \citep{mininni2009Finite}.
For the inverse cascade ranges of H40 and M40, $H_M(k)\sim k^{-8/3}\sim E_b(k)/k$, implying a fully helical magnetic field.
In the evolutionary process given in SM \citep{supplemental_material}, $H_M(k)\sim k^{-10/3}$ in the inverse cascade range \citep{muller2012Inverse,linkmann2017Triad}. 
Furthermore, the comparison of M40 and H40 shows that Hall effects enhance large-scale magnetic energy and helicity by nearly tenfold.
This implies Hall effects could strongly affect large-scale magnetic fields, which is the main subject of this letter.

\par 
Then, the filtering approach \citep{camporeale2018Coherent,manzini2022Local} is used for the details of cascades.
The fluxes for the magnetic energy ($\widetilde{ \Pi}_{Eb1}$ and $\widetilde{ \Pi}_{Eb2}$) and magnetic helicity ($\widetilde{ \Pi}_{HM1}$ and $\widetilde{ \Pi}_{HM2}$) are considered:
\begin{subequations}
  \allowdisplaybreaks
  \begin{align}
    \widetilde \Pi_{Eb1}&=-\epsilon_{ijk}\tilde j_i \tau_{jk}({\bf v},{\bf{b} }),\\
    \widetilde \Pi_{Eb2}&=\epsilon_{ijk}d_I\tilde j_i \tau_{jk}({\bf j},{\bf b}),\\
    \widetilde \Pi_{HM1}&=-2\epsilon_{ijk}\tilde b_i \tau_{jk}({\bf v},{\bf b}),\label{equ:fflux_HM1}\\
    \widetilde \Pi_{HM2}&=2\epsilon_{ijk}d_I\tilde b_i \tau_{jk}({\bf j},{\bf b}),
  \end{align}
  \label{equ:fflux}%
\end{subequations}
where ${\bf \tau}_{ij}({\bf p},{\bf q}  )=\widetilde{{  p_i q_j}}-\widetilde{{p}_i }\widetilde{{q}_j }$ for arbitrary vectors  $\bf p$ and $\bf q$, and the superscript $\widetilde{\cdot}$ represents the filtered quantities.
$\widetilde\Pi_{Eb1}$ and $\widetilde\Pi_{HM1}$ both appear in MHD and HMHD turbulence, while $\widetilde\Pi_{Eb2}$ and $\widetilde\Pi_{HM2}$ only appear in HMHD turbulence.
The overall magnetic energy $\widetilde\Pi_{Eb}=\widetilde\Pi_{Eb1}+\widetilde\Pi_{Eb2}$, and the overall magnetic helicity $\widetilde\Pi_{HM}=\widetilde\Pi_{HM1}+\widetilde\Pi_{HM2}$.
For filtering approaches, different filters lead to consistent qualitative results \citep{yan2020Dual,kuzzay2019Local}, and the fourth-order Butterworth filter  \citep{camporeale2018Coherent} is  used in this study.
\begin{figure}[tb]
  \centering
  \includegraphics[width=\linewidth]{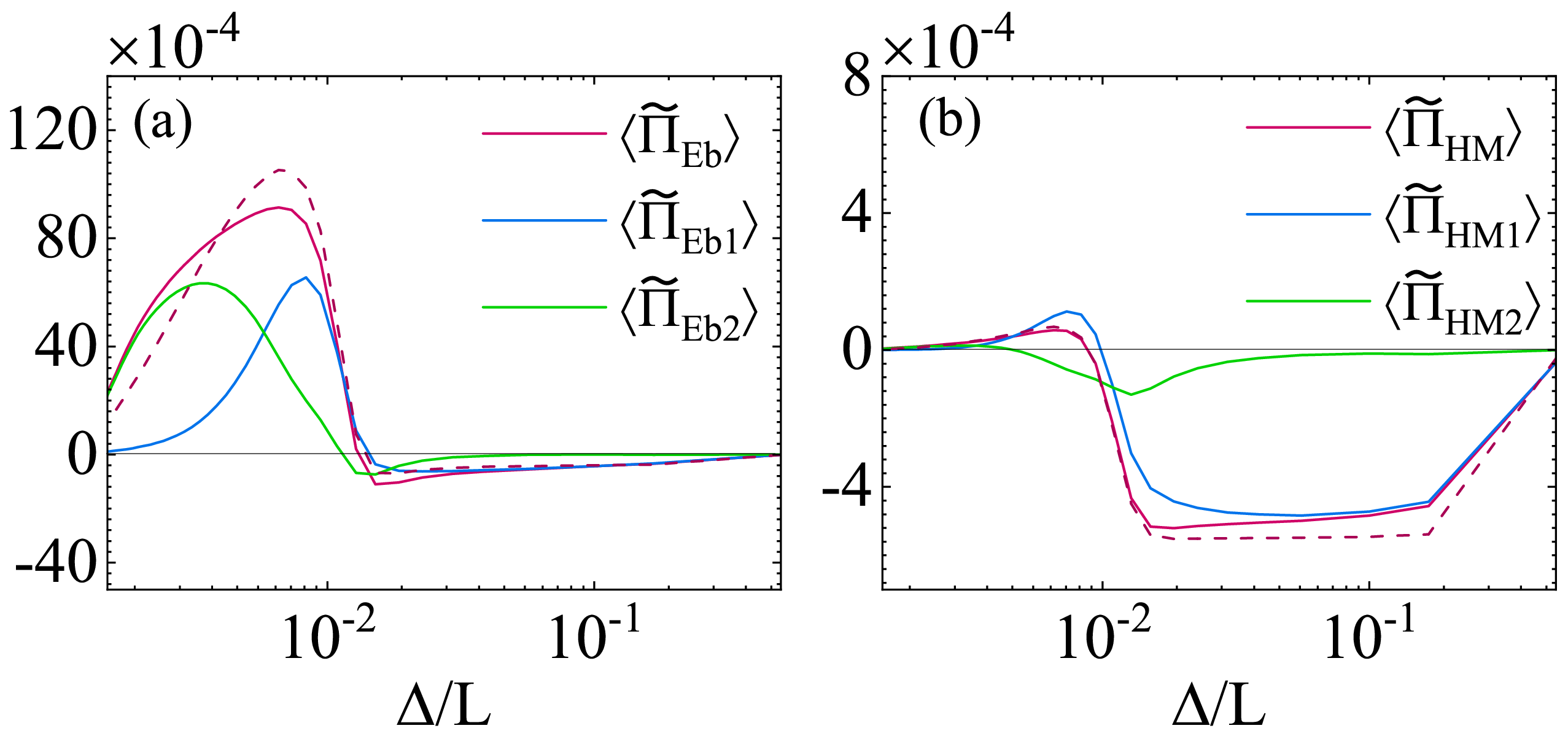}
  \caption{
    Fluxes versus the filter width $\Delta$, where $L=2\pi$.
  (a) Magnetic energy fluxes;
  (b) Magnetic helicity fluxes.
  Solid lines: H40; dashed lines: M40.
  }
  \label{fig:flux_filtering}
\end{figure}
FIG.~\ref{fig:flux_filtering}  (a) shows the  magnetic energy fluxes of  H40 and M40, where $\left\langle \cdot \right\rangle $ represents the ensemble average.
As shown, most magnetic energy cascades forward towards small scales. Weak inverse cascades also appear and have been identified as the $\alpha$-effects \citep{muller2012Inverse,linkmann2017Triad,pouquet2019Helicity}.
FIG.~\ref{fig:flux_filtering} (b) shows the magnetic helicity fluxes.
$\left\langle \widetilde{\Pi}_{HM1} \right\rangle $ is the dominant term, and most magnetic helicity cascades inversely towards large scales.
Hall effects slightly reduce the inverse cascade rates, attributed to the far greater large-scale magnetic helicity and hypo-viscosity.
In summary, Hall effects minimally affect the inverse cascades of magnetic energy and helicity in FIG.~\ref{fig:flux_filtering}, but strongly enhance the large-scale magnetic energy and helicity in FIG.~\ref{fig:spec}.
Therefore, it can be inferred that the enhanced large-scale magnetic field may be related to the inefficiency of cascades.
\par
\begin{figure}[tb]
  \centering
  \includegraphics[width=\linewidth]{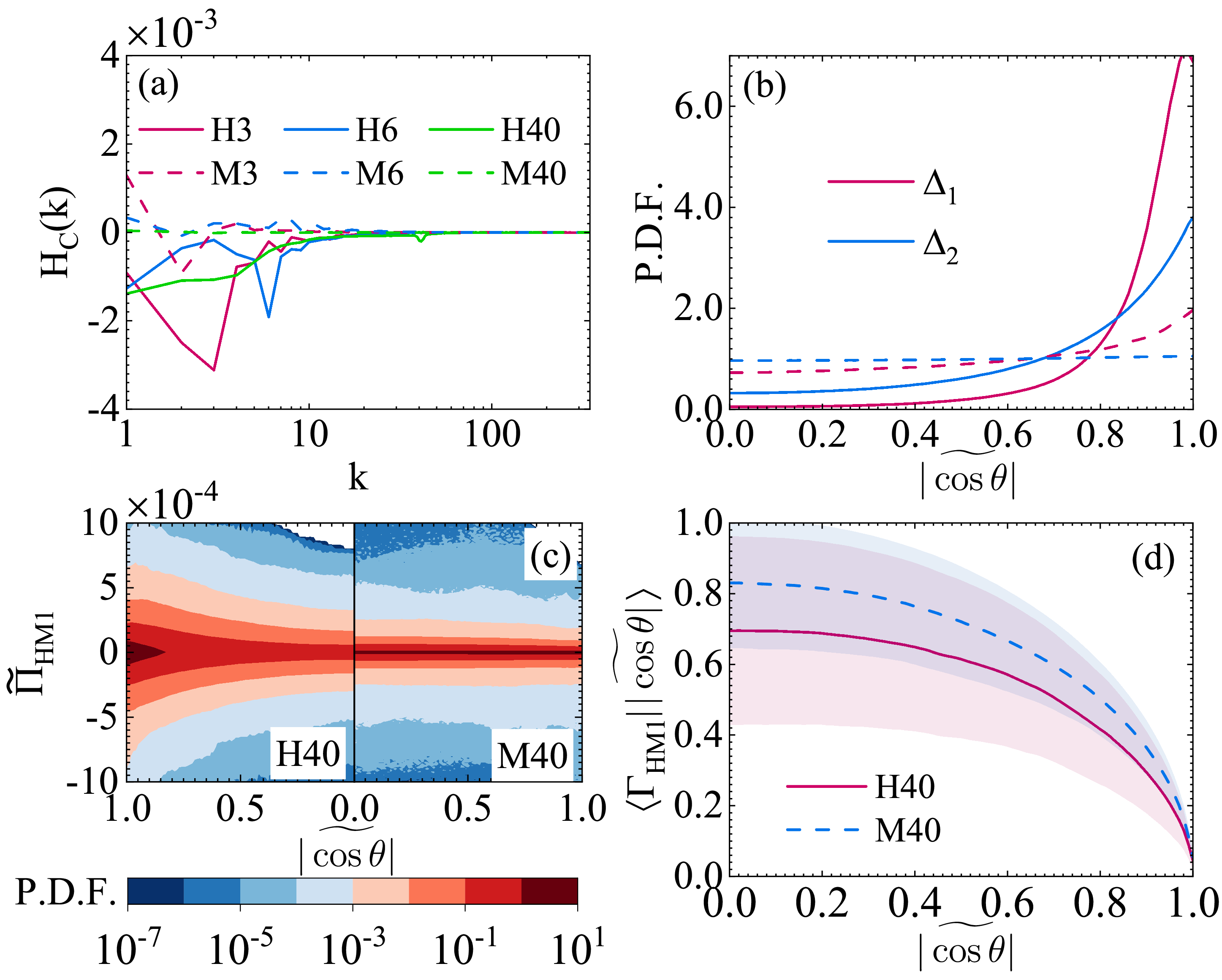}
  \caption{
    Cross helicity distributions and magnetic helicity cascades.  
  (a) $H_C(k)$.
  (b) P.D.F. of $|\widetilde{\cos\theta}|$, with $\Delta_1=0.038L$ and $\Delta_2=0.013L$.
  Solid lines: H40; Dashed lines: M40.
  (c) Joint P.D.F. of  $\widetilde{\Pi}_{HM1}$ and $|\widetilde{\cos\theta}|$ with $\Delta=0.038L$.
  (d) Conditional average of  $\Gamma_{HM1}$ versus $|\widetilde{\cos\theta}|$, with $\Delta=0.038L$.
  }
  \label{fig:efficiency}
\end{figure}

 With large scales governed by the inverse magnetic helicity cascades, the dominant term $\left\langle \widetilde{\Pi}_{HM1} \right\rangle $ is the focus henceforth. 
According to Eq.~(\ref{equ:control}) and (\ref{equ:fflux_HM1}), the cascade is directly related to cross helicity.
FIG.~\ref{fig:efficiency} (a) shows the cross helicity spectra $H_C(k)=\sum_{|{\bf k}|=k}\Re\{ \hat {\bf v}({\bf k})\cdot \hat {\bf b}^\ast({\bf k})\}$.
As shown, $H_C(k)$ is approximately zero for MHD  cases but negative for HMHD  cases.
In fact, for MHD cases, cross helicity is conservative \citep{galtier2016Introduction}.
No cross helicity is injected, making it inherently zero.
In contrast, for HMHD cases, generalized helicity and magnetic helicity are conservative \citep{galtier2016Introduction}. 
The generalized helicity can be decomposed into the sum of magnetic, hydrodynamic and cross helicity weighted by  $d_I$. 
Therefore, the weighted sum of hydrodynamic and cross helicity is conservative.
The hydrodynamic helicity is transformed to cross helicity through the Hall term ($ d_I \nabla \times({\bf j}\times {\bf b})$).
For the physical mechanisms, Hall effects are in fact the ion cyclotron effects on the magnetic field induced by the Lorentz force. 
In MHD cases, only the ion cyclotron effects on the velocity field are considered.
In HMHD cases, the introduction of ion cyclotron effects on the magnetic field, which couples the chiralities of velocity and magnetic fields, produces non-zero cross helicity.
\par Furthermore, the normalized cross helicity can be defined as $\widetilde{\cos \theta }={\widetilde{{\bf v}}\cdot \widetilde{{\bf b}}}/[{|\widetilde{{\bf v}}||\widetilde{{\bf b}}|}]$,
which evaluates the angle between velocity and the magnetic fields. 
The normalized cross helicity directly affects the cascade rates through  the term $\nabla\times ({\bf v}\times {\bf b})$ in Eq.~(\ref{equ:control}) and $\widetilde{\Pi}_{Eb1}$,$\widetilde{\Pi}_{HM1}$ in Eq.~(\ref{equ:fflux}).
As $|\widetilde{\cos \theta }|$ increases, the velocity and magnetic fields align with each other, which could reduce the efficiency  of $\widetilde{\Pi}_{Eb1}$,$\widetilde{\Pi}_{HM1}$.
FIG.~\ref{fig:efficiency} (b) shows the probability distribution functions (P.D.F.) of $|\widetilde{\cos \theta }|$ with two filter widths $\Delta_1=0.038L$ and $\Delta_2=0.013L$.
As the filter width increases or the Hall effects are introduced, $|\widetilde{\cos \theta }|$ centers around 1.0, indicating enhanced alignments between $\bf \widetilde{v} $ and $\bf \widetilde{b}$.
FIG.~\ref{fig:efficiency} (c) compares the joint P.D.F. of $\widetilde{\Pi}_{HM1}$ and $\widetilde{\cos \theta}$ between H40 and M40 with the filter width $\Delta=0.038L$.
For M40, $\widetilde{\Pi}_{HM1}$ are almost independent with $|\widetilde{\cos \theta}|$.
In contrast, as Hall effects are introduced (H40), $\widetilde{\Pi}_{HM1}$ primarily concentrates on the region with significant $|\widetilde{\cos \theta}|$, indicating a potential decrease in efficiency.
For a quantitative estimation,  a direct definition of efficiency $\Gamma_{HM1}$ is introduced as
\begin{equation}
  \begin{aligned}
    \Gamma_{HM1}=
    \frac{|\widetilde{\Pi}_{HM1}| }{2|\widetilde{\bf b}| |\epsilon_{ijk}\tau_{jk}({\bf v},{\bf b}){\bf e}_i|}\frac{|\widetilde{\bf v}\times\widetilde{\bf b}|}{|\widetilde{\bf v}|\,|\widetilde{\bf b}|},
  \end{aligned}
\end{equation}
where ${\bf e}_i$ is the unit vector on the $i$-th direction, the first fraction evaluates the efficiency loss induced by the angle between the two vectors $\widetilde{\bf b}$ and $\epsilon_{ijk}\tau_{jk}({\bf v},{\bf b}){\bf e}_i$, and the second fraction evaluates the direct efficiency loss induced by the angle between $\widetilde{\bf v}$ and $\widetilde{\bf b}$.
FIG.~\ref{fig:efficiency} (d) shows the conditional average $\left\langle \Gamma_{HM1}| |\widetilde{\cos \theta}| \right\rangle $  with  $\Delta=0.038L$.
$\Gamma_{HM1}$ is inversely proportional to $|\widetilde{\cos \theta}|$. 
In addition, H40 has a relatively lower efficiency than M40.
Specifically, when $\Delta=0.038L$, Hall effects reduce  the averaged efficiency $\left\langle \Gamma_{HM1} \right\rangle $ by 53.2\%.
\par In fact, both hydrodynamic helicity in hydrodynamic turbulence \citep{chen2003Joint,yan2020Dual} and cross helicity in HMHD turbulence could affect  cascade efficiencies.
It could be of interest to investigate the disparities between the effects induced by the two helicity.
Generally, hydrodynamic or cross helicity has two sources: geometric and phase alignments \citep{milanese2021Dynamic}.
Taking the cross helicity as an example,  geometric alignment means an increase in $|\widetilde{\cos \theta}|$, while phase alignment denotes a transformation   from negative $\widetilde{\cos \theta}$ to positive $\widetilde{\cos \theta}$.
\citet{milanese2021Dynamic} found that in hydrodynamic turbulence, hydrodynamic helicity cascades forward with dynamic phase alignments, while no geometry alignment is detected.
In comparison, FIG.~\ref{fig:efficiency} (b) reveals that geometric alignments are the key processes in  HMHD turbulence.
The two different alignments lead to different effects on cascades.
As shown in FIG.~\ref{fig:efficiency} (b-d), the geometric alignment induced by Hall effects strongly reduces the cascade efficiency. 
\begin{figure}[b]
  \centering
  \includegraphics[width=\linewidth]{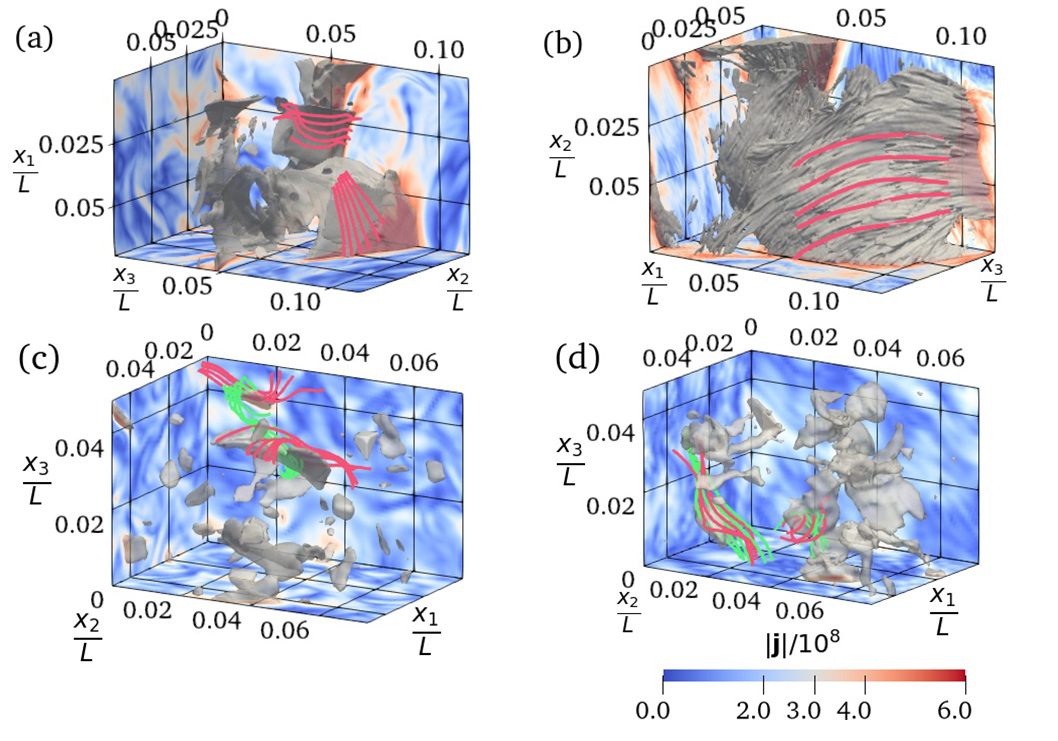}
  \caption{
    Structures with $|{\bf j}|>2.2 \,\textrm{mean}\{|{\bf j}|\}$.
    The background slice shows the contour of $|{\bf j}|$.
    (a) M3; (b) H3; (c) M40; (d) H40.
    Pink lines: magnetic lines;
    Green lines: streamlines.
  }
  \label{fig:current}
\end{figure}
\par FIG.~\ref{fig:current} shows the current density $|{\bf j}|$, where the pink and green lines give the magnetic lines and streamlines, respectively.
The comparison of FIG.~\ref{fig:current} (a) M3 and (b) H3 reveals that when the forcing scale is significantly larger than  $d_I$, Hall effects  induce  filaments along the magnetic field \citep{miura2009Hall}, attributed to the ion cyclotron and whistler modes \citep{banerjee2016Chiral,meyrand2012Spontaneous} and the transverse instabilities \citep{dreher2005Formation} in HMHD.
The comparison of FIG.~\ref{fig:current} (c) M40 and (d) H40 shows that as the forcing scale is close to $d_I$, current sheets are totally broken by these filaments along the magnetic field.
The current density distribution of H40 is irregular and coarse.
Moreover, alignments between the streamlines and magnetic lines are more pronounced in H40 than in M40.
\par Then,  experimental evidence is necessary to verify our numerical findings.
Since only single-point time series of solar wind can be obtained by spacecraft, vKH relations are needed \citep{politano1998Dynamical}.
Exact vKH relations for the magnetic helicity in isotropic HMHD turbulence are derived here, where the bidirectional transfer model  \citep{xie2019Thirdorder} is applied.
After lengthy derivations in SM \citep{supplemental_material}, the first  vKH relation for the magnetic helicity can be written as
\begin{equation}
  \allowdisplaybreaks
  \begin{small}
  \label{equ:VMH1}
  \begin{aligned}
  V_{HM1}(r)&=\frac{r}{2}\langle\delta (U_3b_2-U_2b_3)\delta b_L\rangle\\
    &=\frac{1}{3}\varepsilon_{HMI}\,r-\sum_{K\in \{k_f\}}\varepsilon_{HM}(K) \frac{\sin Kr-Kr\cos Kr}{K^3 r^2},\\
    &=\left\{
    \begin{aligned}
      -\varepsilon_{HMF}\, r/3, r k_f\ll 1,\\
      \varepsilon_{HMI}\, r/3, r k_f\gg 1,\\
    \end{aligned}  
      \right.
  \end{aligned}
\end{small}
\end{equation}
where $\delta q_i({\bf r})=q_i({\bf x}')-q_i({\bf x})$ 
is the two-point increment for an arbitrary vector ${\bf q}$;
${\bf r= x'-x}$ is the displacement between two positions $\bf x$ and $\bf x'$ with a distance $r=|{\bf r}|$;
the subscript $L$ refers to the longitudinal direction (along $\bf r$), and the subscripts 2 and 3 refer to the two remaining transverse directions;
$\varepsilon_{HMI} $ and  $\varepsilon_{HMF}$ are the inverse and forward cascade rates,  $\varepsilon_{HM}(K)$ is the injection rate at the forcing wavenumber $K\in \{k_f\}$.
This equation is the isotropic form of the results derived by \citet{banerjee2016Chiral}.
Using the isotropic condition, $V_{HM1}(r)$  is equivalent to
\begin{equation}
  \label{equ:VMH2}
  \begin{aligned}
    V_{HM2}=2\left\langle \delta (v_2b_L-v_Lb_2) a_2^\ast \right\rangle-{d_I}\left\langle \delta (b_Lb_i) b_i^\ast \right\rangle,
  \end{aligned}
\end{equation}
where $q_i^\ast({\bf r})=[q_i({\bf x})+q_i({\bf x}')]/2$ for an arbitrary vector $\bf q$.
\begin{figure}[b]
  \centering
  \includegraphics[width=\linewidth]{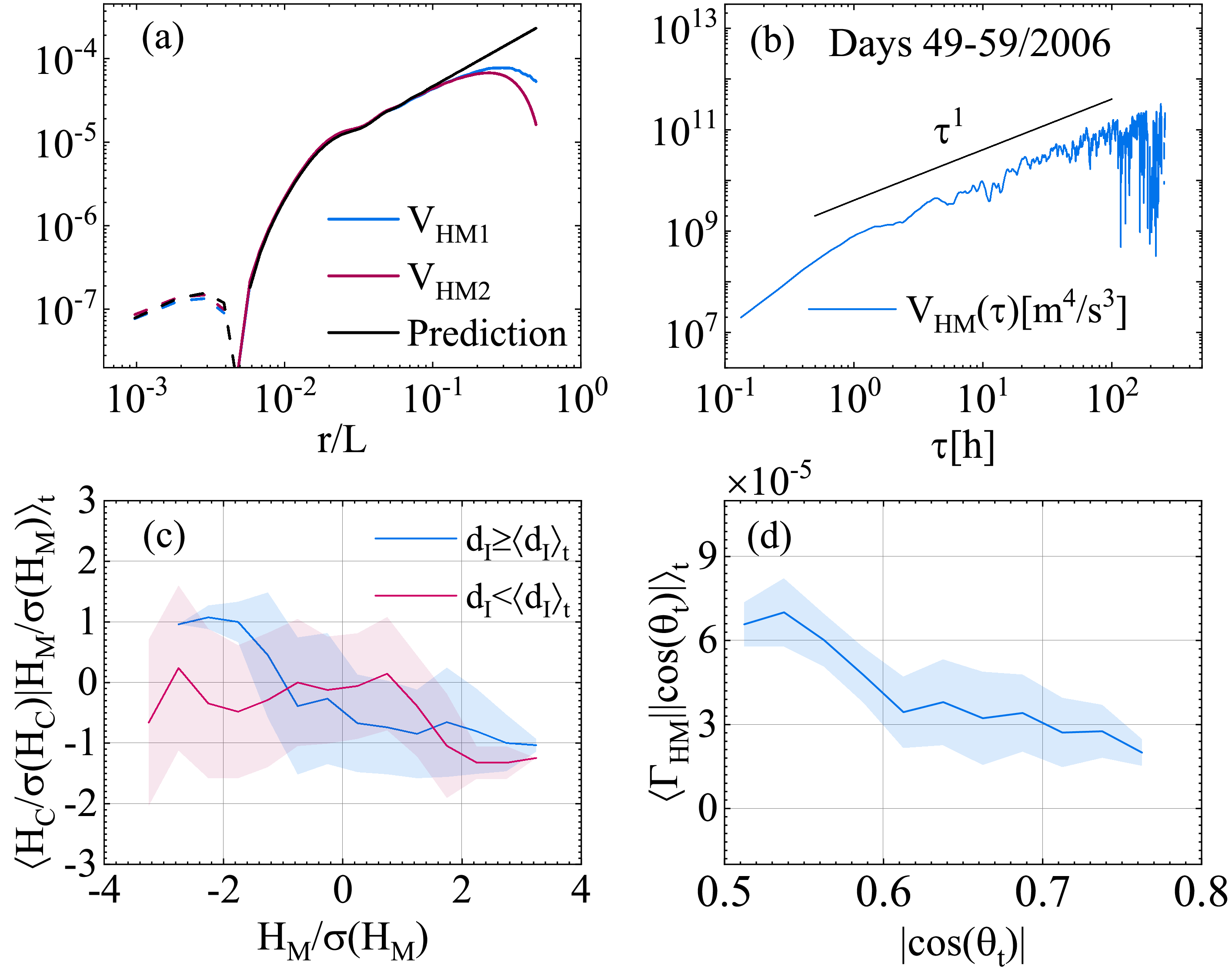}
  \caption{
    Evidence of solar wind data.
    (a) Numerical verifications of $V_{HM1}$ and $V_{HM2}$.
    Solid lines: positive values; dashed lines: negative values.
    (b) Example of the linear scaling  in solar wind.
    (c) Conditional average of the cross helicity versus the magnetic helicity.
    (d) The normalized cascade rates versus the normalized cross helicity. 
  }
  \label{fig:experimental_evidence}
\end{figure}
FIG.~\ref{fig:experimental_evidence} (a) verifies the vKH relations using the results of H40, where black curves are predictions in Eq.~(\ref{equ:VMH1}) with $\varepsilon_{HMI}=4.4\times 10^{-4}$ and $\varepsilon_{HMF}=1.0\times 10^{-4}$. 
The equivalent cascade rates are estimated through the fluxes over corresponding wavenumber ranges  in SM \citep{supplemental_material}.
As shown, $V_{HM1}$ and $V_{HM2}$ are almost the same.
At the scale $r/L < 0.2$, they both fit well with the predictions. 
In contrast, at the scale $r/L\gtrapprox 0.2$, the results deviate from the prediction, attributed to the simple inertial range assumption \citep{xie2019Thirdorder}. 
To further verify the vKH relations,  8 minute averaged time series for solar wind during day 209/1993 - day 253/1994, day 212/1995 - day 224/1996 and day 20/2006 - day 23/2007 of the Ulysses spacecraft are evaluated \citep{balogh1995Heliospheric,smith1995Ulysses}. 
The three periods are very close to solar activity minima, and the heliocentric distance ranges from 2.5 to 4.5 AU \citep{marino2012Occurrence}.
Using the Taylor hypothesis \citep{taylor1938Spectrum}, the vKH relation for magnetic helicity 
$V_{HM}(\tau)= \left\langle  -V_L \tau\delta (v_3b_2-v_2b_3)\delta b_L/2-    d_I\delta(b_Lb_i) b_i^\ast \right\rangle _t=-\varepsilon_{HM}V_L\tau/3$,
where $V_L$ is the averaged longitudinal velocity, $\tau$ is the time lag,  $\left\langle \cdot \right\rangle _t$ represents the time average, $b_i$ is  normalized by $(4\pi \rho)^{-1/2}$, and $\rho$ is the mass density.
FIG.~\ref{fig:experimental_evidence} (b) gives one example of linear scaling related to $V_{HM1}(\tau)$, using data of days 49-59 in 2006.
As shown,  $V_{HM1}(\tau)$ is linear among nearly two decades with $\varepsilon_{HM}=-8.9\times 10^{11} \textrm{J}\cdot \textrm{m/(kg}\cdot \textrm{s})$. 
\par Based on vKH relations, the turbulent space plasma can be recognized, and the magnetic  helicity dissipation rates can be measured. 
Limited by the data resolution, the Hall effects on the inverse cascades cannot be directly recognized.
Therefore, we decompose the numerical finding into two parts: the cross helicity generation, and the cascade efficiency suppression by cross helicity.
Notably, in the actual solar wind, the cross helicity could also originate from other effects.
In FIG.~\ref{fig:experimental_evidence} (c), the conditional average of the cross helicity ($H_C$) versus the magnetic helicity ($H_M$) is evaluated.
Averaged over every 11 days, $H_M$ and $H_C$ at the timescale $\delta t\in (8 \,\textrm{min},4 \,\textrm{h})$ are obtained by the Fourier expansion   \citep{brandenburg2011Scale,matthaeus1982Measurement}.
The results at other timescales are consistent with our main findings and are given in SM \citep{supplemental_material}.
As shown in FIG.~\ref{fig:experimental_evidence} (c), when the Hall effects are relatively important ($d_I\ge \left\langle d_I \right\rangle _t$),
$H_C$ is strongly anticorrelated with $H_M$.
In contrast, when $d_I< \left\langle d_I \right\rangle _t$, $H_C$ is relatively independent with $H_M$, except for the extreme event ($H_M/\sigma(H_M)\gtrapprox 2$).
Even if the data with an 8-min resolution is insufficient to resolve the Hall effects, the remarkable anticorrelation between $H_C$ and $H_M$ when $d_I\ge \left\langle d_I \right\rangle _t$ imply that Hall effects and magnetic helicity could induce non-zero cross helicity. 
FIG.~\ref{fig:experimental_evidence} (d) shows the conditional average of the normalized cascade rates $\varGamma_{HM}=|\varepsilon_{HM}|/[|\delta {\bf v}||\delta {\bf b}|^2V_L \tau]$ versus the normalized cross helicity $|\cos \theta_t|=|\delta {\bf v}\cdot \delta {\bf b}|/|\delta {\bf v}||\delta {\bf b}|$, where the distance ${ r}$ is given by the middle scales with linear scaling.
As shown, the normalized cascade rates are anticorrelated with $|\cos \theta_t|$,  consistent with our numerical findings.
\par In this letter, we perform six DNSs with grid points up to $2048^3$ to investigate Hall effects on helical MHD turbulence. 
We find that  Hall effects could induce strong cross helicity by geometric alignments, which then strongly reduces inverse cascade efficiencies and leads to the accumulation of magnetic energy and helicity at large scales.
The process is associated with the breaking of current sheets by filaments along magnetic fields.
Then, we derive two vKH relations  and verify them using the numerical and Ulysses data.
The numerical findings about the cross helicity generation and the cascade efficiency suppression are then separately confirmed by the solar wind results.
In the heliosphere, partial extreme solar wind events are generated by the eruption of the solar corona, where magnetic helicity and Hall effects both matter. 
Generally, one may believe that  Hall effects are negligible in large-scale solar wind \citep{marino2023Scaling}.
However, this letter gives a possible mechanism in solar wind turbulence that  Hall effects at small scales could strongly affect  large-scale magnetic fields through cross helicity.
\begin{acknowledgments}
  This work was supported by the National Key Research and Development Program of China (Grant Nos. 2019YFA0405300 and 2020YFA0711800) and NSFC Projects (Grant Nos. 91852203, 12072349, 92052102 and 12272006).
  The authors thank the Ulysses project for providing the solar wind data used in this study.
\end{acknowledgments}

%
  
\end{document}